\documentclass[
    ,final            
  ]
  {aipproc}

\layoutstyle{6x9}

\begin{document}

\newcommand{\half}{\frac{1}{2}}
\newcommand{\e}{\mathrm{e}}
\newcommand{\ii}{\mathrm{i}}
\newcommand{\dd}{\mathrm{d}}
\newcommand{\rmd}{\mathrm{d}} 
\newcommand{\tr}{\mathrm{tr}}
\newcommand{\T}{\mathrm{T}}

\newcommand{\deint}[2]{\dd^{#1}\! #2\;}
\newcommand{\de}{\partial}
\newcommand{\dev}{\vec{\de}}
\newcommand{\bra}{\langle}
\newcommand{\ket}{\rangle}

\newcommand{\HIGS}{HI$\gamma$S\xspace}
\newcommand{\threeHe}{${}^3$He\xspace}

\newcommand{\Q}{P}
\newcommand{\w}{\ensuremath{\omega}}
\newcommand{\wcm}{\ensuremath{\omega_\mathrm{cm}}}
\newcommand{\wlab}{\ensuremath{\omega_\mathrm{lab}}}
\newcommand{\wlabout}{\ensuremath{\omega_\mathrm{lab}^\prime}}
\newcommand{\wBreit}{\ensuremath{\omega_\mathrm{Breit}}}
\newcommand{\thetacm}{\ensuremath{\theta_\mathrm{cm}}}
\newcommand{\thetalab}{\ensuremath{\theta_\mathrm{lab}}}
\newcommand{\thetaBreit}{\ensuremath{\theta_\mathrm{Breit}}}
\newcommand{\alphae}{\ensuremath{\alpha_{E1}}}
\newcommand{\betam}{\ensuremath{\beta_{M1}}}
\newcommand{\gammaee}{\ensuremath{\gamma_{E1E1}}}
\newcommand{\gammamm}{\ensuremath{\gamma_{M1M1}}}
\newcommand{\gammaem}{\ensuremath{\gamma_{E1M2}}}
\newcommand{\gammame}{\ensuremath{\gamma_{M1E2}}}
\newcommand{\gammazero}{\ensuremath{\gamma_{0}}}
\newcommand{\gammapi}{\ensuremath{\gamma_{\pi}}}
\newcommand{\alphaep}{\ensuremath{\alpha_{E1}^{(\mathrm{p})}}}
\newcommand{\betamp}{\ensuremath{\beta_{M1}^{(\mathrm{p})}}}
\newcommand{\gammaeep}{\ensuremath{\gamma_{E1E1}^{(\mathrm{p})}}}
\newcommand{\gammammp}{\ensuremath{\gamma_{M1M1}^{(\mathrm{p})}}}
\newcommand{\gammaemp}{\ensuremath{\gamma_{E1M2}^{(\mathrm{p})}}}
\newcommand{\gammamep}{\ensuremath{\gamma_{M1E2}^{(\mathrm{p})}}}
\newcommand{\gammazerop}{\ensuremath{\gamma_{0}^{(\mathrm{p})}}}
\newcommand{\gammapip}{\ensuremath{\gamma_{\pi}^{(\mathrm{p})}}}
\newcommand{\alphaen}{\ensuremath{\alpha_{E1}^{(\mathrm{n})}}}
\newcommand{\betamn}{\ensuremath{\beta_{M1}^{(\mathrm{n})}}}
\newcommand{\gammaeen}{\ensuremath{\gamma_{E1E1}^{(\mathrm{n})}}}
\newcommand{\gammammn}{\ensuremath{\gamma_{M1M1}^{(\mathrm{n})}}}
\newcommand{\gammaemn}{\ensuremath{\gamma_{E1M2}^{(\mathrm{n})}}}
\newcommand{\gammamen}{\ensuremath{\gamma_{M1E2}^{(\mathrm{n})}}}
\newcommand{\gammazeron}{\ensuremath{\gamma_{0}^{(\mathrm{n})}}}
\newcommand{\gammapin}{\ensuremath{\gamma_{\pi}^{(\mathrm{n})}}}
\newcommand{\alphaes}{\ensuremath{\alpha_{E1}^{(\mathrm{s})}}}
\newcommand{\betams}{\ensuremath{\beta_{M1}^{(\mathrm{s})}}}
\newcommand{\gammaees}{\ensuremath{\gamma_{E1E1}^{(\mathrm{s})}}}
\newcommand{\gammamms}{\ensuremath{\gamma_{M1M1}^{(\mathrm{s})}}}
\newcommand{\gammaems}{\ensuremath{\gamma_{E1M2}^{(\mathrm{s})}}}
\newcommand{\gammames}{\ensuremath{\gamma_{M1E2}^{(\mathrm{s})}}}
\newcommand{\gammazeros}{\ensuremath{\gamma_0^{(\mathrm{s})}}}
\newcommand{\gammapis}{\ensuremath{\gamma_\pi^{(\mathrm{s})}}}
\newcommand{\alphaev}{\ensuremath{\alpha_{E1}^{(\mathrm{v})}}}
\newcommand{\betamv}{\ensuremath{\beta_{M1}^{(\mathrm{v})}}}
\newcommand{\gammaeev}{\ensuremath{\gamma_{E1E1}^{(\mathrm{v})}}}
\newcommand{\gammammv}{\ensuremath{\gamma_{M1M1}^{(\mathrm{v})}}}
\newcommand{\gammaemv}{\ensuremath{\gamma_{E1M2}^{(\mathrm{v})}}}
\newcommand{\gammamev}{\ensuremath{\gamma_{M1E2}^{(\mathrm{v})}}}
\newcommand{\gammazerov}{\ensuremath{\gamma_{0}^{(\mathrm{v})}}}
\newcommand{\gammapiv}{\ensuremath{\gamma_{\pi}^{(\mathrm{v})}}}

\newcommand{\MN}{\ensuremath{M_\mathrm{N}}} 
\newcommand{\Mp}{\ensuremath{M_\mathrm{p}}} 
\newcommand{\Mn}{\ensuremath{M_\mathrm{n}}} 
\newcommand{\Md}{\ensuremath{M_\mathrm{d}}} 
\newcommand{\MHe}{\ensuremath{M_\text{\threeHe}}} 
\newcommand{\MDelta}{\ensuremath{M_\Delta}} 
\newcommand{\mpi}{\ensuremath{m_\pi}}     
\newcommand{\fpi}{\ensuremath{f_\pi}}
\newcommand{\wpi}{\ensuremath{\omega_\pi}}
\newcommand{\MeV}{\ensuremath{\mathrm{MeV}}}
\newcommand{\fm}{\ensuremath{\mathrm{fm}}}
\newcommand{\ChiEFT}{$\chi$EFT\xspace}
\newcommand{\EFTNoPion}{EFT($\slashed{\pi}$)\xspace}
\newcommand{\NoPion}{\ensuremath{\slashed{\pi}}}
\newcommand{\LambdaNoPion}{\ensuremath{\Lambda_\slashed{\pi}}}
\newcommand{\QNoPion}{\ensuremath{Q_\slashed{\pi}}}

\newcommand{\NXLO}[1]{N\ensuremath{{}^{#1}}LO\xspace}

\newcommand{\kv}{\vec{k}}
\newcommand{\pv}{\vec{p}}
\newcommand{\qv}{\vec{q}}
\newcommand{\wave}[3]{\ensuremath{{}^{#1}\mathrm{#2}_{#3}}\xspace}
\newcommand{\oneS}{\wave{1}{S}{0}}
\newcommand{\threeS}{\wave{3}{S}{1}}

\renewcommand{\Re}{\ensuremath{\mathrm{Re}}}
\renewcommand{\Im}{\ensuremath{\mathrm{Im}}}

\renewcommand{\deg}{\ensuremath{^\circ}}
\newcommand{\OdL}{Olmos de Le\'on}


\newcommand{\calA}{\mathcal{A}}\newcommand{\calB}{\mathcal{B}}\newcommand{\calC}{\mathcal{C}}
\newcommand{\calD}{\mathcal{D}}\newcommand{\calE}{\mathcal{E}}\newcommand{\calF}{\mathcal{F}}
\newcommand{\calG}{\mathcal{G}}\newcommand{\calH}{\mathcal{H}}\newcommand{\calI}{\mathcal{I}}
\newcommand{\calJ}{\mathcal{J}}\newcommand{\calK}{\mathcal{K}}\newcommand{\calL}{\mathcal{L}}
\newcommand{\calM}{\mathcal{M}}\newcommand{\calN}{\mathcal{N}}\newcommand{\calO}{\mathcal{O}}
\newcommand{\calP}{\mathcal{P}}\newcommand{\calQ}{\mathcal{Q}}\newcommand{\calR}{\mathcal{R}}
\newcommand{\calS}{\mathcal{S}}\newcommand{\calT}{\mathcal{T}}\newcommand{\calU}{\mathcal{U}}
\newcommand{\calV}{\mathcal{V}}\newcommand{\calW}{\mathcal{W}}\newcommand{\calX}{\mathcal{X}}
\newcommand{\calY}{\mathcal{Y}}\newcommand{\calZ}{\mathcal{Z}}

\newcommand{\ga}{g_{\scriptscriptstyle A}}
\newcommand{\gpiNN}{g_{\pi{\scriptscriptstyle\text{NN}}}}

\newcommand{\lambdachi}{\Lambda_\chi}
\newcommand{\ChPT}{\ensuremath{\chi \mathrm{PT}}}
\newcommand{\alphaEM}{\alpha_{\rm EM}}
\newcommand{\GeV}{\ensuremath{\mathrm{GeV}}}
\newcommand{\NsqLO}{\ensuremath{\mathrm{N^2LO}}}
\newcommand{\gE}{\ensuremath{g_{\scriptscriptstyle E}}}
\newcommand{\gM}{\ensuremath{g_{\scriptscriptstyle M}}}
\newcommand{\be}{\begin{equation}}
\newcommand{\ee}{\end{equation}}
\newcommand{\bea}{\begin{eqnarray}}
\newcommand{\eea}{\end{eqnarray}}
\newcommand{\nn}{\nonumber}
\newcommand{\gaprox}{${\raisebox{-.6ex}{{$\stackrel{\textstyle >}{\sim}$}}}$}
\newcommand{\simge}{\hspace*{0.2em}\raisebox{0.5ex}{$>$}
     \hspace{-0.8em}\raisebox{-0.3em}{$\sim$}\hspace*{0.2em}}
\newcommand{\simle}{\hspace*{0.2em}\raisebox{0.5ex}{$<$}
     \hspace{-0.8em}\raisebox{-0.3em}{$\sim$}\hspace*{0.2em}}
\newcommand{\order}[1]{{\cal O}(#1)}

\title 
      [EFT analysis of $\gamma$p and $\gamma$d scattering]
      {Using EFT to analyze low-energy Compton scattering from protons and light nuclei}

\author{Daniel R. Phillips}{
  address={Institute for Nuclear and Particle Physics and Department of Physics and Astronomy, Ohio University, Athens, OH 45701, USA },
  email={phillid1@ohio.edu}
}

\iftrue
\author{Judith McGovern}{
  address={Theoretical Physics Group, School of Physics and Astronomy, University of Manchester, Manchester, M13 9PL, United Kingdom},
  email={judith.mcgovern@man.ac.uk},
}

\author{Harald W. Grie\ss hammer}{
  address={Institute for Nuclear Studies, Department of Physics, The George Washington University, Washington, DC 20052, USA},
  email={hgrie@gwu.edu},
}
\fi

\copyrightyear  {2012}

\begin{abstract}
We discuss the application of an effective field theory (EFT) which incorporates the chiral symmetry of QCD to Compton scattering from the proton and deuteron. We describe the chiral EFT analysis of the $\gamma$p scattering database presented in our recent review~\cite{Gr12}, which gives:
\begin{center}
$
  \alphaep=10.5\pm0.5(\rm{stat})\pm0.8{\rm(theory)}\;\;,\;\;
  \betamp = 2.7\pm0.5(\rm{stat})\pm0.8(\rm{theory}),
$
\end{center}
for the electric and magnetic dipole polarizability of the proton.
We also summarize Ref.~\cite{Gr12}'s chiral EFT analysis of the world data on coherent Compton scattering from deuterium, which yields:
\begin{center}
$
  \alphaes=10.5\pm 2.0(\rm{stat})\pm0.8(\rm{theory})\;\;,\;\;
  \betams=3.6\pm 1.0(\rm{stat})\pm0.8(\rm{theory}).
$
\end{center}
\end{abstract}


\date{\today}

\maketitle

\section{Theoretical Background}

Experiments to measure Compton scattering from the proton and deuteron are presently being pursued at a number of facilities around the world, including MAMI (Mainz)~\cite{Hornidge,Miskimen}, HI$\gamma$S at TUNL~\cite{We09}, and MAX-Lab at Lund~\cite{Fe08}.  Chiral effective field theory ($\chi$EFT) is one of the main theoretical techniques used to analyze these experiments. $\chi$EFT generates the most general Compton amplitude that is consistent with electromagnetic gauge invariance, the pattern of chiral-symmetry breaking in QCD, and Lorentz covariance, to any given order in the small parameter $P \equiv \{\omega,m_\pi\}/\Lambda$, with $\omega$ the photon energy, $m_\pi$ the pion mass and $\Lambda$ the breakdown scale of the theory. 

The pioneering calculations of $\gamma$p scattering in $\chi$EFT~\cite{Be91,Be95} were performed in a theory with only nucleons and pions as explicit degrees of freedom. This reduces the breakdown scale $\Lambda$ to the energy at which the $\Delta(1232)$ is excited, i.e. $M_\Delta - M_N \approx 300$ MeV. 
 In Refs.~\cite{Be91,Be95} nucleon scalar dipole polarizabilities are predicted to be, at $\order{P^3}$,
 \begin{eqnarray}
   \alphae&=&10\betam=\frac{10\alphaEM\ga^2}{192\pi\mpi\fpi^2}=12.6\times 10^{-4}~{\rm fm}^3.
   \label{eq:op3preds}
  \end{eqnarray}
The corresponding cross sections agree well with experiment up to at least $\omega \sim m_\pi$, but do not capture the rise of the data towards the $\Delta(1232)$ peak. In this variant of $\chi$EFT even an $\order{P^4}$ calculation cannot describe the data at backward angles once $\omega \simge 180$ MeV~\cite{McG01,Be03,Be05}.

The inclusion of the Delta as an active degree of freedom in the theory is therefore essential if the full power of the world's Compton data to shed light on fundamental hadron-structure parameters, e.g. polarizabilities, is to be realized. With the $\Delta(1232)$ included in $\chi$EFT~\cite{Bu92,He96,Hi04} the ratio $(M_\Delta - M_N)/\Lambda$ becomes one of the theory's expansion parameters. 
Ref.~\cite{PP03} pointed out that $(M_\Delta-M_N)/\Lambda$ is numerically rather similar to the expansion parameter in $\Delta$-less calculations, $m_\pi/(M_\Delta-M_N)$, and denoted both as $\delta$.   In  the ``$\delta$-counting'' adopted in Ref.~\cite{PP03} powers of the electronic charge $e$ are shown explicitly, while Refs~\cite{Be91,Be95} counted $e\sim P$.  Thus the Thomson amplitude is $\order{P^2}\sim\order{e^2\delta^0}$ and structure effects start with $\pi$N loops at $\order{P^3}\sim\order{e^2\delta^2}$ in the low-energy region. Further, since $M_\Delta - M_N \sim \delta$, whereas $m_\pi \sim \delta^2$, $\pi \Delta$ loops are suppressed by an additional power of $\delta$, and do not enter the amplitude until $\order{e^2 \delta^3}$. 

The Delta-pole graph has a special role in $\delta$-counting: it too is $\order{e^2 \delta^3}$ for $\omega \sim m_\pi$, but it becomes enhanced in the region $\omega \sim M_\Delta - M_N$, because of proximity to the Delta's on-shell point~\cite{PP03}. In this domain 
the effects that generate the resonance's finite width must be resummed. The dominant $\gamma$p scattering mechanism in this ``medium-energy'' region is then the excitation of a dressed $\Delta(1232)$ by the magnetic transition from the nucleon state, followed by de-excitation via the same M1 transition. This effect occurs at $\order{e^2 \delta^{-1}}$. At $\order{e^2 \delta^0}$ the E2 $N \rightarrow \Delta(1232)$ transition must also be considered. Further discussion of $\chi$EFT, and this power counting, can be found in Ref.~\cite{Gr12}. 

Eq.~(\ref{eq:op3preds}) is also the prediction for the \emph{neutron} polarizabilities in (Delta-less) $\chi$EFT at $\order{P^3}$. To access $\alphaen$ and $\betamn$ experimentally a nuclear target is required. The deuteron is the simplest nucleus, and an accurate description of its structure is obtained when $\chi$EFT is applied to the NN problem (see, e.g., Refs.~\cite{EMG05,Ph06}). 
$\gamma$d scattering was first calculated in ($\Delta$-less) $\chi$EFT in Ref.~\cite{Be99} at $\order{P^3}$, where it was demonstrated that there are large isoscalar $\gamma NN \rightarrow \gamma NN$ mechanisms (``exchange currents'') which are crucial to obtaining reasonable agreement with the data. This calculation was extended to higher orders, and augmented with $\Delta(1232)$ degrees of freedom, in Ref.~\cite{Be03,Be05,Hi04B}. However, it was not until the work of Ref.~\cite{Hi05} that a $\chi$EFT treatment of Compton scattering from deuterium which respected the Thomson limit for the $\gamma$d amplitude was formulated. This computation also included $\Delta(1232)$ degrees of freedom, and demonstrated the ability of $\chi$EFT to describe deuterium Compton data from threshold to $\omega \simge 100$ MeV.

\section{A new analysis of $\gamma$\lowercase{p} scattering in $\chi$EFT}

The calculation of the $\gamma$p differential cross section presented here includes  the nucleon Born graph and the $t$-channel $\pi^0$ pole graph (both calculated covariantly).  The  $\Delta$-pole graphs ($s$- and $u$-channel) are dealt with as described in Refs.~\cite{Gr12,PP03,LP09}---covariantly and with a finite width stemming from $\pi$N loops.  Compton $\pi$N and $\pi \Delta$ loop graphs are also added, so the amplitude includes effects which are of leading or next-to-leading order throughout the kinematic region $0 \leq \omega_{\rm lab} \simle 350$ MeV (apart from  the loop correction to the $\gamma$N$\Delta$ vertices), and all effects up to next-to-next-to-leading order---$\order{e^2\delta^3}$---in the low-energy region $\omega \sim m_\pi$. In addition,
we include the contact interactions which encode the short-distance ($r \ll 1/m_\pi$) contributions to the scalar polarizabilities. Strictly speaking these do not occur until $O(e^2 \delta^4)$ in the low-energy domain, but they are necessary for an accurate description of $\gamma$p data in HB$\chi$EFT (c.f. the case where pion loops are calculated relativistically~\cite{LP09,Le12}). The coefficients of these contact terms are fit to the $\gamma$p database, which is equivalent to fitting $\alphaep$ and $\betamp$.

The parameters used in the $\pi$N sector take standard values (see Ref.~\cite{Gr12}).  The $\Delta(1232)$ parameters $M_\Delta - M_N= 293$~MeV and $g_{\pi N\Delta} =1.425$ are obtained from the Breit-Wigner peak and width, the latter via the relativistic formula. We adopt  
$b_2/b_1=-0.34$ for the ratio of E2 and M1 couplings (c.f. the $\chi$EFT study of pion photoproduction~\cite{PV06}).

Three EFT parameters---$b_1$, $\alphaep$, and $\betamp$---remain to be determined. They are fit to the 
$\gamma$p data base discussed extensively in Ref.~\cite{Gr12}. For the reasons explained there, in the low-energy region
we include data from Refs.~\cite{Hy59,Go60,Pu67,Ba74,Fe91,Zi92,Ha93,MacG95,OdeL}. We float the normalization of each of these data sets within the quoted normalization uncertainty. In the medium-energy region the data sets of Refs.~\cite{Ha93,Blanpied} are in significant disagreement with those from MAMI (most notably 
Refs.~\cite{Wolf,Ca02}) and a consistent fit cannot be obtained. We have chosen to use the MAMI data for our fits in this region~\cite{Gr12}.

\begin{figure}[!t]
    \includegraphics*[width=\linewidth]{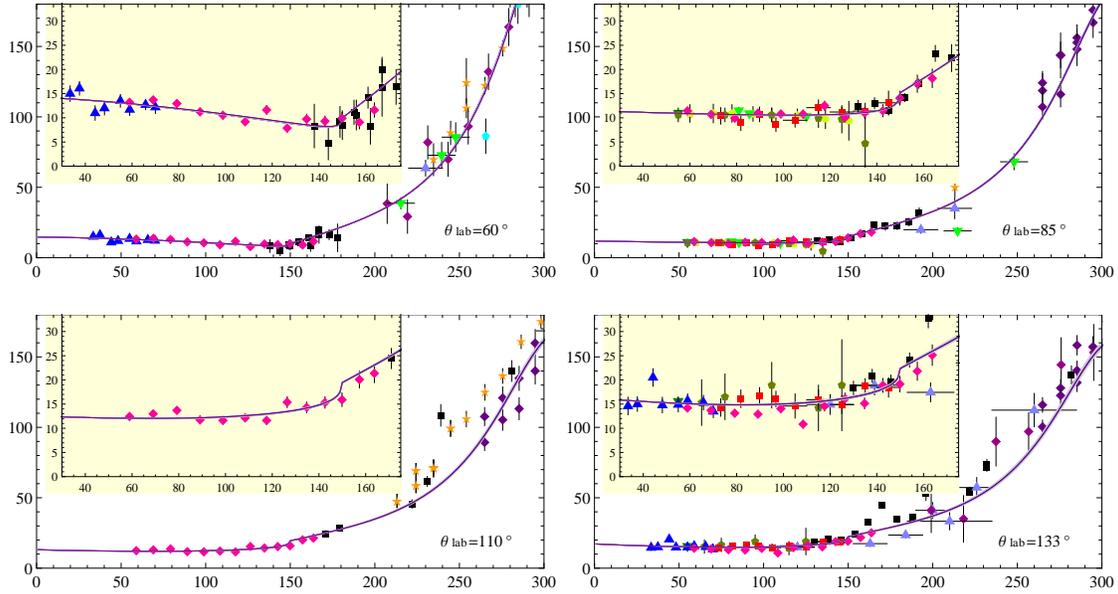}
    \caption {Comparison of our $\chi$EFT result for $\gamma$p scattering with
      data.  The lab cross section in nb/sr is shown in 10$^\circ$ bins for $\theta_{\rm lab}$
       as a function of $\omega_{\rm lab}$ in MeV, with insets
      showing the fit region. The grey band shows the variation within the statistical error of the one-parameter fit. Adapted from Ref.~\cite{Gr12}, where the legend for experimental data can be found.}
    \label{fig:gammap}
\end{figure}

Our strategy is to determine the $\gamma$N$\Delta$ M1
coupling $b_1$ by considering the MAMI data for $\wlab=$200--325~MeV, then fit $\alphaep$ and $\betamp$ to the low-energy data up to 170 MeV, and iterate until
convergence is reached. The $\chi^2/{\rm d.o.f.}$ of the low-energy Hallin data is hard to accept, and so we prefer to quote our best results without them. We 
then obtain a solution
with a  $\chi^2/{\rm d.o.f.}= 106.1/124$. This ``modified $\calO(e^2\delta^3)$''  fit yields $b_1=3.66\pm 0.03$ and the values of $\alphaep$ and $\betamp$ quoted in the abstract. Since these sum to a value consistent with the Baldin sum-rule constraint
$\alphaep+\betamp = 13.8\pm0.4$~\cite{OdeL}, one can impose this relation to find $\alphaep-\betamp=7.7\pm 0.6$
with $\chi^2/{\rm d.o.f.}=106.5/125$, an unchanged $b_1$, and the cross sections
displayed in Fig.~\ref{fig:gammap}.
These fits are stable against reasonable variations in the procedure~\cite{Gr12}, and 
 agree with data \cite{Wolf,Ca02,Be60,Ba66,Ge76,PH97,Wi99} well beyond the region in which the free parameters are determined (see Fig.~\ref{fig:gammap}).

\section{A new analysis of $\gamma$\lowercase{d} scattering in $\chi$EFT}

The $\chi$EFT treatment of $\gamma$d scattering developed in Ref.~\cite{Hi05} is valid
from threshold to $\wlab \simge 100$ MeV, and represents a
complete (modified) $\calO(e^2\delta^3)$ calculation in both the two- and
one-nucleon sectors. It has the added virtue that the dependence of cross sections on the choice 
of deuteron wave function is 
$< 1$\% (c.f. Ref.~\cite{Be05}). We now employ the $\chi$EFT deuteron wave function at NNLO~\cite{EMG05} within this theory to extract $\alphaes$ and $\betams$ from the 
$\gamma$d elastic data of Refs.~\cite{Lu94,Ho99,Lu03}. 

\begin{figure}[!htb]
    \includegraphics*[width=0.5\linewidth]{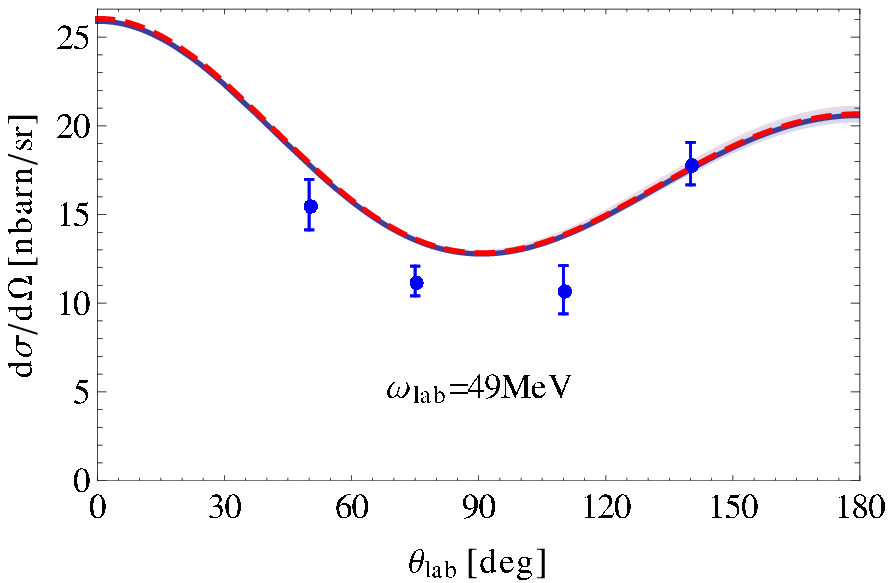}
    \includegraphics*[width=0.5\linewidth]{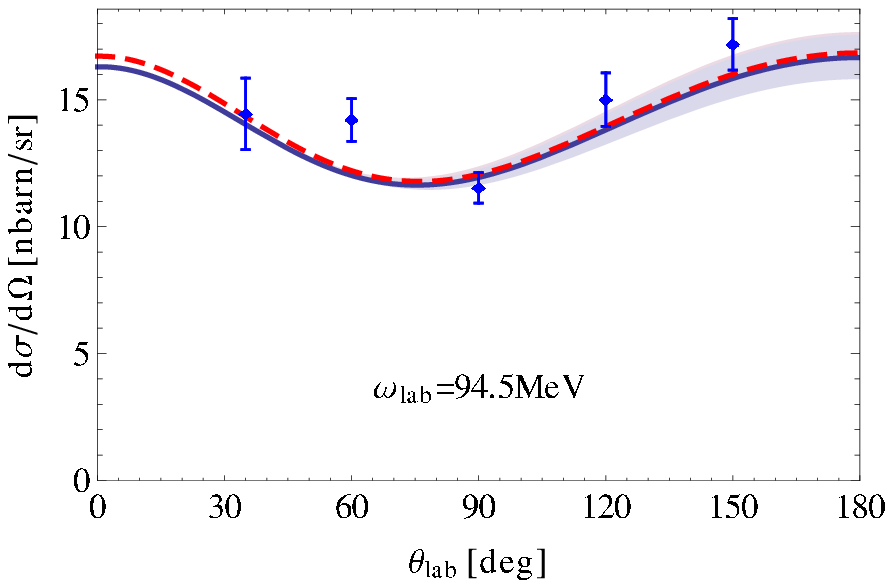}
    \caption{$\gamma$d cross sections at 49 and 94.5 MeV in the two-parameter (dashed) and
      one-parameter (solid) determinations of the isoscalar spin-independent
      dipole polarisabilities. Bands: statistical error of the Baldin constrained
      fit. Data at 49 (94.5) MeV from Ref.~\cite{Lu94} (\cite{Ho99}). Adapted from Ref.~\cite{Gr12}. \label{fig:dfit}}
\end{figure}

The fit to these isoscalar combinations of scalar dipole polarizabilities yields the results quoted in the abstract,
with $\chi^2/{\rm d.o.f.}=24.3/24$ (see Fig.~\ref{fig:dfit}). In contrast to the proton case, this is a consistent data base: each experiment contributes roughly 
equally to the $\chi^2$, and the extracted polarizabilities are largely insensitive to the elimination of any one data set.
The isoscalar polarizabilities we obtain are close to the
proton ones, so isovector effects in $\alphae$ and $\betam$ are
small, as predicted by $\chi$EFT at $\order{P^3}$.
We also used the (isoscalar) Baldin constraint to reduce statistical uncertainties in a
one-parameter fit, and obtained very similar results (see Fig.~\ref{fig:dfit}). For further details, see Ref.~\cite{Gr12}.


\begin{theacknowledgments}
D.~R.~P. thanks the organizers of CIPANP 2012 for a stimulating meeting in a great location, and,
in particular, Vladimir Pascalutsa and Michel Guidal for co-ordinating the ``Nucleon structure'' 
session.
 This work was supported in part by
UK Science and Technology Facilities Council grants ST/F012047/1, ST/J000159/1
(JMcG) and ST/F006861/1 (DRP), by the US Department of Energy under grants
DE-FG02-95ER-40907 (HWG) and DE-FG02-93ER-40756
(DRP), by the US National Science Foundation \textsc{Career} award PHY-0645498
(HWG), and by University Facilitating Funds of the George Washington
University (HWG).

\end{theacknowledgments}



\bibliographystyle{aipproc}   

\end{document}